\documentclass[twocolumn,showpacs,preprintnumbers,superscriptaddress,amsmath,amssymb]{revtex4}
\usepackage{amsfonts}
\usepackage{amsmath}
\usepackage{amssymb}
\usepackage{graphicx}%
\usepackage{footmisc}
\usepackage{bm}
\usepackage{braket}
\usepackage{CJK}
\usepackage{hyperref}
\newcommand{\abs}[1]{\lvert#1\rvert}
\newcommand{\norm}[1]{\lVert#1\rVert}
\newcommand{\Tr}{\text{Tr}}
\begin{document}
\begin{CJK*}{UTF8}{gbsn}
\title{Lieb-Robinson Bound at Finite Temperature}

\author{Zhiqiang Huang (黄志强)}
\email{hzq@wipm.ac.cn}
\affiliation{Wuhan Institute of Physics and Mathematics, Chinese Academy of Sciences, Wuhan 430071, China}
\affiliation{University of the Chinese Academy of Sciences, Beijing 100049, China}

\author{Xiao-Kan Guo (郭肖侃)}
\email{xkguo@wipm.ac.cn}
\affiliation{Wuhan Institute of Physics and Mathematics, Chinese Academy of Sciences, Wuhan 430071, China}
\affiliation{University of the Chinese Academy of Sciences, Beijing 100049, China}

\date{\today}

\begin{abstract}
The Lieb-Robinson bound shows that the speed of propagating information in a nonrelativistic quantum lattice system is bounded by a finite velocity, which entails the clustering of correlations. In this paper, we extend the Lieb-Robinson bound to quantum systems at finite temperature by calculating the dynamical correlation function at nonzero temperature for systems whose interactions are respectively short-range,  exponentially-decaying and long-range. We introduce a simple way of counting the clusters in a cluster expansion by using the combinatoric generating functions of graphs. Limitations and possible applications of the obtained bound are also discussed.
\end{abstract}


\maketitle
\end{CJK*}

\section{Introduction}
In relativistic physics, the speed of light defines a light cone for propagating particles, and hence sets the causality of the theory. For nonrelativistic quantum spin systems, the Lieb-Robinson bound \cite{LR72} defines a similar upper bound on the speed of  signals/information propagating on a lattice. For a given lattice $\Gamma$ and  a quantum spin system defined on $\Gamma$ with Hamiltonian $H$, if two  observables represented by self-adjoint  operators $A,B$ are supported respectively on the sublattices $X\subset\Gamma,Y\subset\Gamma$, the Lieb-Robinson bound says that for the time evolution of $A$, i.e. $\tau_t(A)=e^{iHt}Ae^{-iHt}$, the operator norm of the commutator $[\tau_t(A),B]$ can be bounded as
\begin{equation}\label{1}
\norm{[\tau_t(A),B]}\leqslant c\exp\{-a(d(X,Y)-v_{\text{LR}}\abs{t})\}
\end{equation}
where $a,c$ are positive constants and $d(X,Y)$ is the distance between $X$ and $Y$. $d(X,Y)$ is $0$ if $X$ and $Y$ overlap and otherwise equal to the size of the smallest subset of edges that connects $X$ and $Y$. Here the constant $v_{\text{LR}}$ is the Lieb-Robinson velocity, which sets the upper bound of the propagation of information (say, perturbations) in this system.

The causality defined by a bounded speed of information propagation (plus unitarity) implies the localizability principle, even in the quantum case (for which see e.g. \cite{BGNP01}), that all physical properties can
be attributed to local mechanisms. Using the Lieb-Robinson bound, one can show the exponential decay of correlations \cite{NS06,HK06}, which manifests the localizability of the lattice system. More explicitly, if we denote by $\braket{A}_\omega$ the  expectation value of the observable $A$ in a state  $\omega$, then 
 \begin{equation}\label{2}
\abs{\braket{AB}_{\omega }-\braket{A}_{\omega }\braket{B}_{\omega }}\leqslant K\norm{A}\norm{B}\min(\abs{X},\abs{Y}) e^{-ad(X,Y)}
\end{equation}
where $K,a$ are positive constants and $\abs{X}$ is the number of sites in the support $X$. This inequality \eqref{2} holds for gapped ground states in systems  with respectively short-range and exponentially decaying interactions. 

The exponential decay  or clustering of correlations is an example of how the Lieb-Robinson bound helps us in understanding the structures and dynamics of nonrelativistic quantum systems. There are many other applications of the Lieb-Robinson bound, for which one is referred to the review articles \cite{NS10,KGE14}. Among these applications, however, the Lieb-Robinson bound for quantum systems at finite temperature has not been  considered particularly. A moment of reflection  shows some possible reasons for this and some relevant progresses: On the one hand,  the time evolution of an observable at finite temperature could be stochastic, which might not be as simple as being unitary for a closed system. In \cite{Pou10} a first step has been made  that the Lieb-Robinson bound holds for general Markovian quantum evolutions. 
On the other hand, from the standpoint of quantum statistical mechanics, since the Lieb-Robinson bound  depends only on the Hamiltonian of the system, one could expect that this bound shoud hold at certain equilibrium temperature. Indeed, for two (fermionic) operators $A,B$ at finite temperature in a system with short-ranged interactions, the correlation at finite temperature  also decreases exponentially with respect to temperature \cite{Has04},
\begin{equation}\label{3}
\abs{\braket{AB}_\beta}\leqslant c\norm{A}\norm{B}\exp\{-d(X,Y)/(v\beta)\}
\end{equation}
where $\braket{AB}_\beta=Z^{-1}\text{Tr}[ABe^{-\beta H}]$ is the thermal correlation function and $v$ is a constant velocity. But this inequality \eqref{3} is not dynamical and hence unable to manifest the causal propagation of information. Besides,  the Lieb-Robinson bound enters the proof of \eqref{3} in a  peculiar way that the time parameter is finally integrated out.
An alternative proof of the exponential decay of  equal-time correlations at finite temperature is recently provided by Kliesch {\it et al.} \cite{KGKRE14}. Using the method of cluster expansion, they prove the exponential decay of the generalized thermal covariance for operators in a system with short-range interactions,
\begin{equation}\label{4}
\abs{\text{cov}^\tau_\beta(A,B)}\leqslant\frac{4a \norm{A}_\infty \norm{B}_\infty}{(1-e^{-1/\xi(\beta)})\ln3}e^{-d(X,Y)/\xi(\beta)}
\end{equation}
where the generalized covariance $\text{cov}^\tau_\beta(A,B)=\text{Tr}(\rho^\tau A\rho^{1-\tau}B)-\text{Tr}(\rho A)\text{Tr}(\rho B)$ with $\rho=Z^{-1}e^{-\beta H}$ and $\tau\in[0,1]$, $\norm{\cdot}_\infty$ is the Schatten $\infty$-norm (which is just the operator norm $\norm{\cdot}$), and $\xi(\beta)$ is a function depends on $\beta$. From \eqref{4} one can find a critical temperature for localization in this system. 

Although the exponential decays in \eqref{3} and \eqref{4} hold at a nonzero temperature, they do not manifests the time-evolution of observables. The study of dynamical correlation functions at finite temperature, however, is a common practice in statistical mechanics and thermal field theories. For example, the exponential clustering of dynamical correlations of the Gibbs states on a one-dimensional lattice has been proved long ago in \cite{A69}. But the discussions there are restricted to the one-dimensional lattice with finite range interactions and the bounded speed of information propagation is still difficult to know.
In this paper, we will combine the iterating integral representations in proving the Lieb-Robinson bound \cite{HK06} and the cluster expansions \cite{KGKRE14} to give a Lieb-Robinson bound at finite temperature in the systems with respectively power-law decaying, exponentially decaying and short-range interactions. We prove that the exponential decay or power-law decaying of correlations  always persists at high enough temperature for short enough time. For instance, for systems with exponentially decaying interactions we have a bound on the Schatten 1-norm of the dynamical thermal correlation function at finite temperature,
\begin{widetext}
\begin{equation}\label{5}
	\norm{\text{Tr}([A(t), B] \rho(\beta))}_1 \leqslant \frac{4C_0 Z(\beta)|X||Y|\left\|A\right\|_\infty \left\|B\right\|_\infty}{(1+ d(A,B))^\epsilon\ln3} \left(\frac{f(\beta \lambda_0)}{1-f(\beta \lambda_0)}\right)  {\exp^{-\mu d(A,B)+2\lambda_0 p_0 |t|} },
\end{equation}
\begin{figure}[t]
\centering
\includegraphics[width=0.8\textwidth]{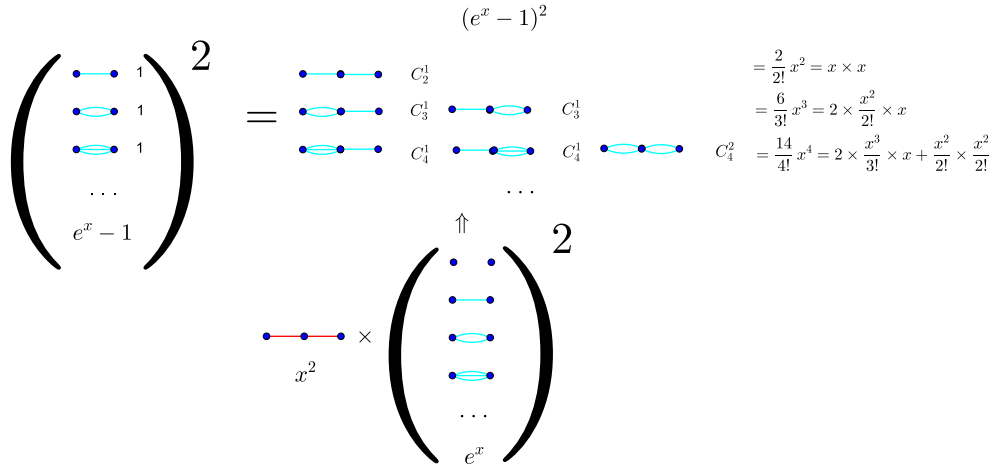}
\caption{
The generating function for two-site graphs with nonempty bonds is $\mathcal{G}_s(x)=\sum_{k=1}^{\infty}  x^{k}/k!=e^x-1$, and the two-site generating function allowing empty bond is $\mathcal{G}'_s=\sum_{k=0}^{\infty}  x^{k}/k!=e^x$. These two types of graphs are depicted in the big brackets.
The generating function for graphs with $m$ different (nonempty) bonds is $\mathcal{G}_m(x)=(e^x-1)^m$. The graphs of $\mathcal{G}_m(x)$ can be generated from those of $\mathcal{G}_s(x)$ by squaring. 
The lower part of figure shows that  adding $\mathcal{G}'_s$ to the skeleton will lead to  a larger generating function.}
\label{fig.c}
\end{figure}
\end{widetext}
where $A(t)=\tau_t(A)$, $\epsilon,\mu,\lambda_0,p_0,C_0$ are positive constants, $\rho(\beta)=e^{-\beta H}/Z$, $f(\beta)$ is a function of $\beta$ whose explicit form can be found in Sec. \ref{CCC}, \eqref{nnbf1}, and $d(A,B)$ is the distance between their supports $d(X,Y)$. In systems with other interactions the results are similar to \eqref{5}. In comparison with \eqref{1} as well as \eqref{3},\eqref{4}, we see that it gives both the critical temperatures and the Lieb-Robinson velocities. And when the temperature is fixed, the result reduces to the original Lieb-Robinson bound (times a constant factor).

In the next section, we first introduce a cluster expansion of the partition function defined on a graph by exploiting the generating functions for graphs, which greatly simplifies the proofs involving cluster expansions as those in \cite{KGKRE14}. In Sec. \ref{sec3}, we prove the main results of this paper, that is, the Lieb-Robinson bounds at finite temperature for systems with interactions of various type.  And finally in Sec. \ref{dis}, we  discuss some possible applications of the Lieb-Robinson bound at finite temperature and conclude the paper.

\section{Cluster expansion on graphs}\label{II22}
For a general two-diemsional  quantum lattice system, there are a set $V$ of  sites (or vertices)  and a set $E$ of their bonds (or edges). The Hamiltonian $H$ is defined on the tensor product $\otimes _{x\in V}\mathcal{H}_x$ of  finite dimensional Hilbert spaces $\mathcal{H}_x$ at the sites $x\in V$. The interaction is defined by the sum of mutually commuting local Hamiltonians  on the bonds 
\begin{equation}
	H=\sum_{l\in E} h_l=\sum_{x,y\in l\in E}h_{x,y}.
\end{equation}
Here every bond $l\in E$ corresponds to a pair of sites $\{x,y\}\subset V$, so that $h_l$ or $h_{x,y}$ is defined on $\mathcal{H}_x\otimes\mathcal{H}_y$. If we say an $l$ meets the $X\subset E$ as $l\cap X\neq \emptyset$, it means $\{x,y\}\cap V(X)\neq \emptyset$, which further means that the bonds can  meet on sites only. 


Now that the interactions are defined on the bonds, the cluster expansion of partition functions, either in  the real-time case for time evolutions or in the imaginary time case for statistical mechanics, can be expressed in terms of the sequences of bonds, which amounts to the evaluation of the expansion in terms of (sub)graphs,
\begin{equation}\label{7}
	e^{\alpha H}=\sum_{g\in G} \frac{\alpha^{|g|} h(g)}{|g|!},
\end{equation}
where $G$ is the set of all possible graphs, $\abs{g}$ is the number of bonds in $g$, and by factorizability of bonds $h(g)=\prod_{l\in g}h_l$. A graph $g$ here is marked by a sequences of bonds $(l_1,l_2,\dots,l_{|g|})$.  
However, not all the graphs are involved in the concrete physical process. There will be many graphs that are simply unphysical. We can put some constraints to pick out 
the subset $G'\subset G$ that is physical. To this end, if we can bound the interactions by some bond-independent number such as $\left\|h(g)\right\|_\infty=J^{|g|}$, the problem will be translated into finding the generating function $\mathcal{G}(x)$ of the graphs $g$, and setting $x=|\alpha J|$. In view of \eqref{7}, the constraints can be enforced on the form of the generating function such that
\begin{equation}
\mathcal{G}(x)=\sum_k \frac{g_k x^{k}}{k!},\quad x:\text{a parameter}
\end{equation}
with $g_k$, the number of graphs with $k$ bonds, unspecified. 

The generating functions $\mathcal{G}$ are employed to generate different graphical structures from a particular graphical structure. A schematical example is the following: Given a non-connected graphs $\Gamma_1$ with restriction rules $P$ (say, on the possible linking of bonds), let us define a generating function $\mathcal{G}_1(x)$. Similarly, for a connected graph $\Gamma_2$ with the same restriction rules $P$ as $\Gamma_1$, define a generating function $\mathcal{G}_2(x)$. Then we can decompose the non-connected $\Gamma_1$ into many connected $\Gamma_2$'s. In terms of generating functions, this is an algebraic equation $\mathcal{G}_1=\mathfrak{A}(\mathcal{G}_2)$ where $\mathfrak{A}$ denotes some algebraic operations that transform $\mathcal{G}_2$ to $\mathcal{G}_1$. (E.g. in the case of decomposing non-connected to connected graphs $\mathfrak{A}=\exp$ if the connected subgraphs do not overlap each other.)
Let us outline some simple but concrete examples of specifying the $g_k$ and the corresponding generating function for some graphs. For more details, one is referred to \cite{HP73}.
For a single bond, if we repeat it $k$ times with only two fixed sites, it  gives a single kind of graphs $g_0=0,g_{k>0}=1$. So, the generating function for such  graphs is $\mathcal{G}_s(x)=\sum_{k=1}^{\infty}  x^{k}/k!=e^x-1$.
 If we include the empty bond, then $g_{k}=1$, and the generating function is $\mathcal{G}'_s=\sum_{k=0}^{\infty}  x^{k}/k!=e^x$. Similarly, for graphs with $m$ different (nonempty) bonds, each of which can be repeated $k$ times, the generating function is simply $\mathcal{G}_m(x)=(e^x-1)^m$. We thus see that the graphs with $m$ bonds are obtained from the $\mathcal{G}_s(s)$ through exponentials. See also Fig. \ref{fig.c} for illustrations.

There are some graphs that are not easy to count.
We need to neglect some properties of these graphs, so that they become simpler graphs, and then add these properties to finally recover the original graphs. For instance, one can simplify a graph to its skeleton tree, then the recovering process can be viewed as adding new graphs to the tree with some specific rules.
These new graphs can also be enumerated with generating functions.
Therefore, the generating function of the original graphs can be obtained by multiplying the generating function of the simplified graphs and the added graphs. Take the graphs with $m$ bonds and $\mathcal{G}_m$ given above as an example. We can choose the simplified graph as the graph of $m$ bonds without repetition, and then add a bond to it in the following way: (i) The generating function of $m$ bonds without repetition is $\mathcal{G}'_m(x)=x^m$. (ii) When we add a repeated bond, we have $m$ choice, and hence every added bond contributes a factor $mx$. The generating function of all the added graphs is then $\mathcal{G}''_m(x)=e^{m x}$. 
The graphs generated in this way will have some redundancies. With positive $x$, the generating function is therefore larger than the original one $(e^x-1)^m\leq x^me^{m x}$. (See Fig. \ref{fig.c}) But this is useful in bounding the cluster expansion, as we shall show in the following.

We can give a short proof of the locality of temperature with the help of generating functions of graphs (because the key step in \cite{KGKRE14} is the cluster expansion as \eqref{13}). The outline of the proof is the following, which is also the procedure we shall follow in the next section.  In  calculating the covariance, the expansion of the partition function is limited to the graphs that have at least $L$ bonds, where $L$ is the distance  between two local operator $A$ and $B$. The graphs can be divided into several blocks of connected graphs, some of which connect to the operator $A$ and others  not. Those blocks that do not connect to $A$ can be viewed as all possible graphs $G$ subtracting the blocks $\mathcal{B}_A$ connecting to $A$, i.e. $\mathcal{B}_{\bar{A}}=G-\mathcal{B}_A$. Every block can be simplified to a connected graph without repetition, which is  called an animal. The number of animals in the lattice with $m$ bonds are bounded by $(K e)^m$ with $K$ being the valence of each site and the $e$ is just the natural constant. So we can start from animals whose number of bonds is bigger than $L$, and add the repeating bonds to recover the blocks, and glue the blocks to recover the original graphs.

\section{Lieb-Robinson bound at finite temperature }\label{sec3}
In this section, we prove a Lieb-Robinson bound at finite temperature in quantum lattice systems using the cluster expansions. We show how to bound the interactions by some bond-independent number, and then utilize the generating functions of graphs to arrive at a Lieb-Robinson bound at finite temperature.

The key quantity in the proof is the dynamical correlation function at finite temperature $\braket{[A(t),B]}_\beta$, so we first study the  cluster expansions of these functions in Sec.\ref{CCA}. In subsequent subsections,  we prove the bounds for different types of interaction. The exponential clustering of correlations  then manifest themselves in the intermediate steps of this section.

\subsection{Dynamical correlation functions at finite temperature} \label{CCA}
Given a quantum lattice system at finite temperature $T=1/\beta$, we can observe some local operator $O$ through its thermal expectation value $\braket{O}_\beta=Z^{-1}\text{Tr}[O e^{-\beta H}]$. (Notice that from the standpoint of quantum thermodynamics, one needs not to specify the Gibbs canonical state $\rho(\beta)$. A non-canonical state $\rho$ in the pertinent Hilbert space will do. Here we stick to the quantum statistical mechanical approach in analogy to \cite{KGKRE14}.)

When one adds a local perturbation $\Delta$ to the system Hamiltonian $H$, the expectation value we observe will be changed. It is expected that if the perturbation and the observable are separated by a large distance, the observable cannot feel the perturbation at once.  At high enough temperature and for short
enough time, the change in the expectation value can be bounded as 
\begin{align}
		&\abs{\braket{e^{i (H+\Delta) t} Oe^{-i (H+\Delta) t} }_\beta-\braket{e^{i H t} Oe^{-i H t} }_\beta }\nonumber\\
&\leqslant  \int_0^t  \, ds\abs{\braket{ -i  e^{i H t}[e^{iH(t-s)}Oe^{-iH(t-s)},\Delta]e^{-i H t}  }_\beta}\nonumber\\
	&\leqslant\int_0^t  \, ds \abs{\braket{ [e^{iH(t-s)}Oe^{-iH(t-s)},\Delta] }_\beta }+ \mathcal{O}(\Delta^2).
\end{align}
We can see that in order to find the effect of the perturbation, we need to analyze the dynamical thermal correlation function $\braket{[A(t),B]  }_\beta$ where $A(t)=e^{iHt}Ae^{-iHt}$ is the time evolution of the local operator $A$. 

\begin{figure}[t]
\centering
\includegraphics[width=0.5\textwidth]{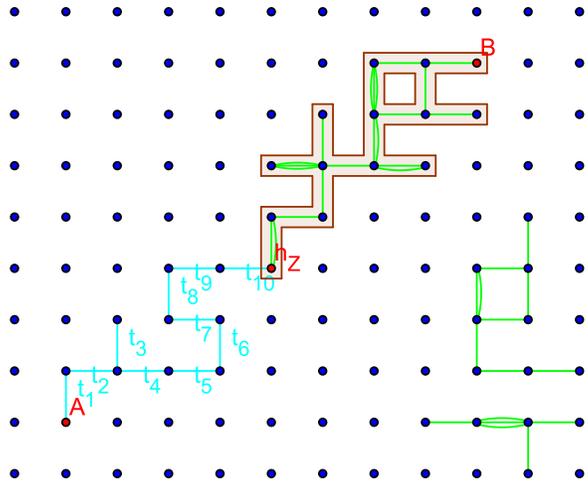}
\caption{The schematic illustration of the cluster expansions of the time evolution part (unshaded) and the partition function (shaded). These two  sequences of bonds  must overlap at some site. The shades are $\mathcal{A}(w)$. There are also graphs (green) that do not connect to $A$, which belongs to $\mathcal{B}_{\bar{A}}$.}
\label{fig.l}
\end{figure}

Here we consider  two local operator $A$ and $B$, supported on $X,Y\subset V$ respectively, in a general quantum lattice system at finite temperature.  (The locality of the operators can be relaxed in such a way that $A$ and $B$ could be non-local operators as the products of local operators supported on regions $X$ and $Y$. Cf. Appendix B of \cite{NO08}. This is because the final bound only depends on the supports $X,Y$.)
We can  give a bound on the absolute value of the dynamical correlation function at finite temperature 
\begin{equation} \label{abf}
  \abs{\braket{[A(t),B]}_\beta} \leqslant\norm{[A(t),B] \rho(\beta) }_1,\quad \rho(\beta)\equiv\frac{e^{-\beta H}}{Z(\beta)}.
\end{equation}
 In order to do so, we need to count graphs appearing in the cluster expansions  of the time evolution operator $e^{-iHt}$ and the Boltzmann factor $e^{-\beta H}$.

Before we perform the cluster expansions, let us make a remark about the connectedness of the graphs in the cluster expansion of $\braket{[A(t),B]}_\beta$.
 Consider the expansion of $A(t)$
 \begin{equation}
 	A(t)=A+it \sum _{Z_{1}\in E} [A,h_{Z_{1}}]+\frac{(it)^2}{2}\sum _{Z_{1},Z_{2}\in E} [[A,h_{Z_{1}}],h_{Z_{2}}]+\dots
 \end{equation}
If the nontrivial expansion of the time evolution part $g(A)$ consists of connected graphs with its contribution in $A(t)$ be labeled as $A_g(t)$, the nontrivial expansion of the partition function must connect to one site of $g(A)$. If not, consider the graphs $h(g)$ that can be separated into two parts $h(g)=h(B)h(A_g)$, where $h(A_g)$ do not connect to $B$ and $h(B)$ do not connect to $g(A)$. In other words, $B\cup h(B)$ have disjoint supports.
The trace can then be factorized into a product of two traces, i.e. $\Tr([A_g(t),B]h(g))=\Tr([A_g(t),h(A_g)][B,h(B)])\propto\Tr([B,h(B)])=0$. Hence, for all those nontrivial contributions, there exists a bond $Z$ connecting $A$ through the expansions of the time evolution part and the partition function to  $B$.

Notice that although their forms look similar, their cluster expansions are very different. The graphs in the nontrivial expansion of the time evolution operator is like the ``paths", whereas the graphs in the nontrivial expansion of the partition function are ``clusters". To see this, let us start the  calculations inspired by the method of \cite{HK06},
\begin{widetext}
\begin{align}   
   \norm{A(t)B\rho(\beta) }_1&= \norm{A(t=0) B \rho(\beta) +\int_0^t \partial_s  A(s) B \rho(\beta )\, ds}_1 \nonumber\\
  &\leqslant \norm{A B \rho(\beta)}_1+\int_0^t  \norm{e^{iHs}[A,H] e^{-iHs}B \rho(\beta) }_1 \, ds\nonumber\\
  &\leqslant\norm{A B \rho(\beta)}_1+\sum _{Z\in E,Z\cap X\neq \emptyset } \int_0^t ( \norm{A(s) h_Z(s)B \rho(\beta) }_1+\norm{ h_Z(s)A(s) B \rho(\beta) }_1) \, ds \nonumber\\
  &\leqslant\norm{A B \rho(\beta)}_1+(\norm{A}_\infty+\norm{h_Z A h_Z^{-1}}_\infty) \sum _{Z\in E,Z\cap X\neq \emptyset } \int_0^t \left\|h_Z(s)B \rho(\beta )\right\|_1 \, ds \nonumber\\
  &\leqslant\norm{A B \rho(\beta)}_1+k\norm{A}_\infty \sum _{Z\in E,Z\cap X\neq \emptyset } \int_0^t \left\|h_Z(s)B \rho(\beta )\right\|_1 \, ds,\label{cff}
\end{align} 
where $\norm{O}_p$ is the Schatten $p$-norm of the operator $O$ and  $k=1+\sigma_{\text{Max}}(h_Z)/\sigma_{\text{Min}}(h_Z)$ with $\sigma_{\text{Max}}(h_Z),\sigma_{\text{Min}}(h_Z)$ being the maximum and minimum singular values of $h_Z$ respectively. The $h_Z$ is a local interaction Hamiltonian defined on the bond $Z$ that does not commute with $A$. Here in the fourth line we have used the property that unitary transformations do not change the operator norm, so that the time dependence $(s)$ is dropped; we have also used the H\"older's inequality $\norm{AB}_1\leqslant\norm{A}_\infty\norm{B}_1$ since $\frac{1}{\infty}+\frac{1}{1}=1$. In the fifth line we have used  $\norm{AB}_\infty\leqslant\norm{A}_\infty\norm{B}_\infty$. Now the 1-norm in the last term of \eqref{cff} resembles the left hand side(LHS) of this inequality, so we can  take the $h(s)$ as the $A(t)$ on LHS and
iterate the derivations. We obtain
\begin{equation} \label{cff1}
   \norm{A(t)B\rho(\beta) }_1\leqslant\norm{A B \rho(\beta)}_1+\left\|A\right\|_\infty \sum_{p:p_1\cap X\neq \emptyset,p_{|p|}\cap Z\neq \emptyset}  \frac{(|k_{\text{Max}} t|)^{|p|+1}}{(|p|+1)!} \left\|h(p)\right\|_\infty\left\|h_Z B \rho(\beta)\right\|_1,
\end{equation}\end{widetext}
where $k_{\text{Max}}$ is the maximum of $k$ among all of the local interactions $h_Z$, the $p$ are sequences of bonds $(p_1,p_2,\dots,p_{|p|})$ that meet sequentially, i.e. $p_1\cap p_2\neq \emptyset,p_2\cap p_3\neq \emptyset,\dots,p_{|p|-1}\cap p_{|p|}\neq \emptyset$,  and $\norm{h(p)}_\infty=\norm{h_{p_1}}_\infty\norm{h_{p_2}}_\infty\dots \norm{h_{p_{|p|}}}_\infty$.  
The coefficient $\abs{t}^{\abs{p}+1}/(\abs{p}+1)!$ in \eqref{cff1} comes from the integral $\int_0^t\int_0^t...\int_0^tdsds_1...ds_{\abs{p}}$. The remaining time-dependent part is an integral over $n+1$ time-parameters $ds_i$ after $n$ iterations, e.g.
\[\int_0^t\int_0^{t}...\int_0^{t}\norm{h_Z(s_n) B \rho(\beta)}_1dsds_1...ds_n.\]
Qualitively, this part will be infinitely small if $t$ is very small, and it will diverge if $t$ is very large. However, it is expected that there would be a maximum $t_{\text{max}}$ of the propagation time $t$ after which two operators are surely correlated. We can therefore take the integrand $\norm{h_Z(t) B \rho(\beta)}_1$ as 1 for $t>t_{\text{max}}$, and there is no physical divergence. In either case, the final result is Eq. \eqref{cff1}.
From Eq. \eqref{cff1}, we can see that each bond must overlap the nearby bonds so that the graphs of these possible sequences are just like ``paths". 

As unitary transformations do not change the Schatten 1-norm of operators, with the equation
\begin{equation}
	\left\| B A(t)\rho(\beta )\right\|_1=\left\|e^{iHt} B(-t) A\rho(\beta ) e^{-iHt}\right\|_1,
\end{equation}
we can see that the bound of $\left\| B A(t)\rho(\beta )\right\|_1$ can be obtained from the bound of $ \norm{A(t)B\rho(\beta) }_1$ in \eqref{cff} by simply exchanging $A$ and $B$ and replacing $t$ with $-t$. 
We can also use the property of KMS states, i.e. $\text{Tr}( B A(t)\rho(\beta) )=\text{Tr}( A(t-i\beta)) B\rho(\beta )$, to give a bound, but it will enlarge the bound indeed.

The nontrivial expansion of the partition function is very different. Consider on a bond $Z$,
\begin{equation}\label{13}
  \left\|h_Z B \rho(\beta)\right\|_1=\frac{1}{Z(\beta)}\left\|h_Z B \sum_{w\in W}\frac{(-\beta)^{|w|}}{|w|!}h(w)\right\|_1,
\end{equation}
where the $w$ are again the sequences of bonds $(w_1,w_2,\dots,w_{|w|})$, and $h(w)=h_{w_1}h_{w_2}\dots h_{w_{|w|}}$. Now the only constraint is that $w$ must contain at least one connected subsequence that connects $Z$ and $B$, and all the connected subsequences connecting to $Z$ must also connect to $B$.  

Such a constraint is not convenient for practical calculations, so we need to divide $w$ into several connected subsequences by adding more constraints and lift them at the final stage. Let us put all of the subsequences that do not connect to $B$ into a single  sequence $w_{B^c}$,  and write other distinct  subsequences (complementary to $w_{B^c}$) as $w_{B_1},w_{B_2},\dots,w_{B_m}$. Since $w_{B_i}$ is  connected, it might contain repeated bonds.  If we subtract the repeated bonds, such a graph without repeated bonds is called an {\it animal}. We denote the animals from sequence $w$ by $\mathcal{A}(w)$. Now we can see that the expansion is in terms of those $w_{b_i}$, the ``clusters" of animals. See Fig. \ref{fig.l} for an illustration.
We can further rewrite Eq.~\eqref{13} in the form of the inclusion-exclusion principle,
\begin{widetext}
\begin{equation}\label{14}
  \left\|h_Z B \rho(\beta)\right\|_1=\frac{1}{Z(\beta)}\left\|h_Z B\sum_{m=1}^{\infty} e^{-\beta H} \frac{1}{m!}\prod_{i=1}^{m} \left(\sum_{w_i}\frac{(-\beta)^{|w_i|}}{|w_i|!}h(w_i) e^{\beta H_{\mathcal{A}(w_i)}}\right)\right\|_1.
\end{equation}

\subsection{Systems with short-range interactions} \label{CCB}

For systems with short-range interactions, the local Hamiltonian $h_l$ is nontrivial only on the nearest neighbors of a site. It is  easier to first  bound the term \eqref{14},
\begin{align} 
  \left\|h_Z B \rho(\beta)\right\|_1=&\frac{1}{Z(\beta)}\left\|h_Z B \sum_{m=1}^{\infty} e^{-\beta H} \frac{1}{m!}\prod_{i=1}^{m} \left(\sum_{w_i}\frac{(-\beta)^{|w_i|}}{|w_i|!}h(w_i) e^{\beta H_{\mathcal{A}(w_i)}}\right)\right\|_1\nonumber\\
  \leqslant&
  \left\|h_Z \right\|_\infty \left\|B \right\|_\infty \sum_{m=1}^{\infty}  \frac{1}{m!}\prod_{i=1}^{m} \left(\sum_{w_i}\frac{(\beta)^{|w_i|}}{|w_i|!} \left\|h(w_i)  e^{\beta H_{\mathcal{A}(w_i)}}\right\|_\infty\right)\label{15}\\
  \leqslant&
  \left\|h_Z \right\|_\infty \left\|B \right\|_\infty \sum_{m=1}^{\infty}  \frac{1}{m!}\prod_{i=1}^{m} \left(\sum_{w_i}\frac{|\beta J|^{|w_i|}}{|w_i|!}e^{|\beta J||\mathcal{A}(w_i)|}\right),\label{16}
\end{align}
where $\abs{J}$ is the largest of $\norm{h_l}_\infty$, and $|\mathcal{A}(w_i)|$ is the number of bonds of $\mathcal{A}(w_i)$. Here in \eqref{15} we have used again the H\"older's inequality together with $\norm{e^{-\beta H}}_1= Z(\beta)$ (since $e^{-\beta H}$ is hermitian and positive definite), and in \eqref{16} we have changed $\norm{h_l}_\infty$ to the largest bond-independent $\abs{J}$. 

With Eq.~\eqref{16}, we can now use the generating functions of graphs $w_{i}$ to give the bounds.
Let us denote by $g_\mathcal{A}$ all the graphs that can be added to animals $\mathcal{A}$ so as to recover those connected graphs. Its  generating function  is $\mathcal{G}_{\mathcal{A}}(x)=(e^x-1)^{|\mathcal{A}|}$. Therefore, we can change the sum over $w_i$ to the sum over animals,
\begin{equation}\label{17}
	\sum_{w_i}\frac{|\beta J|^{|w_i|}}{|w_i|!}e^{|\beta J||\mathcal{A}(w_i)|}=\sum_{\mathcal{A}}\left(\sum_{g_{\mathcal{A}}}\frac{|\beta J|^{|w_i|}}{|w_i|!}\right)e^{|\beta J||\mathcal{A}|}=\sum_{\mathcal{A}}(e^{|\beta J|}-1)^{|\mathcal{A}|}e^{|\beta J||\mathcal{A}|},
\end{equation}
where $\mathcal{A}$ is the set of all possible animals connecting $Z$ to $B$ (not restricted to $w_i$). By denoting the distance between $Z$ and $B$ by $d(Z,B)$, we should require $\abs{\mathcal{A}}\geqslant d(Z,B)$. The number of possible animals is generally bounded by its size $N(|\mathcal{A}|=L)\leq (K e)^L$, where $K$ is the number of nearest neighbors or the valence of a site.  Hence, we can substitute \eqref{17} into \eqref{16} and use the conditions on the number of animals to obtain
\begin{align}
  \left\|h_Z B \rho(\beta)\right\|_1\leqslant &\left\|h_Z \right\|_\infty \left\|B \right\|_\infty \left( \text{exp} \left(\sum_{L=d(Z,B)}^\infty\left(Ke(e^{\abs{\beta J}}-1)e^{\abs{\beta J}}\right)^L \right)  -1  \right)   \nonumber\\
=&\left\|h_Z \right\|_\infty \left\|B \right\|_\infty \left( \text{exp}\left(\frac{b(\beta J)^{d(Z,B)}}{1-b(\beta J)}\right)-1\right),\label{18}
\end{align}
\end{widetext}
where  $b(x)\equiv K e(e^{|x|}-1)e^{|x|}$. Now that $e^x-1\leqslant x(e^{x_0}-1)/x_0,\forall x\in[0,x_0]$ , at a high enough temperature, for example $b(\beta J)^{d(Z,B)}/(1-b(\beta J))\leqslant \text{ln}3 $, we then have the familiar form
\begin{equation}\label{nned}
  \left\|h_Z B \rho(\beta)\right\|_1\leqslant 
  \left\|h_Z \right\|_\infty \left\|B \right\|_\infty \frac{e^{-\mu_1 d(Z,B)}}{1-b(\beta J)}\frac{2}{\ln3}.
\end{equation}
where $\mu_1=-\text{ln}(b(\beta J))$.
Notice that in \eqref{18} mathematically it is required that $\abs{b}<1$, and therefore one can  obtain from $\abs{b}=1$ a universal critical inverse  temperature 
\begin{equation}
	\beta^*=\frac{1}{\abs{J}}\ln\left[(1+\frac{1}{2}\sqrt{1+\frac{4}{K e}})\right].
\end{equation}
If the temperature is higher than this universal critical temperature, according to Eq. \eqref{nned}, the correlation between any observation will be exponentially decaying with respect to the distance, so that there will be no long-range order in the system. Hence,
this critical temperature provides a universal  upper bound on physical critical temperatures like the Curie temperature. 

Let us turn to the calculations about the ``paths" in the expansion of the time evolution operator. Because we have $\left\|h(p)\right\|_\infty \left\|h_Z \right\|_\infty\leq|J|^{|p|+1}$, the bound of term $ \sum_{p:p_1\cap X\neq \emptyset,p_{|p|}\cap Z\neq \emptyset}  \frac{(|k_{\text{Max}} t|)^{|p|+1}}{(|p|+1)!} \left\|h(p)\right\|_\infty \left\|h_Z \right\|_\infty$, where $\norm{h_Z}_\infty$ comes from Eq. \eqref{18}, is just the  generating function of the ``paths" connecting $Z$ to $X$. Different from the ``cluster" of animals, a ``path" itself is a connected graph.
 So, there is no need to dividing the graphs into blocks and subtracting the repeated bonds. Now given $L-1$ bonds, when we add the $L$-th bond to the graph, we have $2K-1$ choices since we have to make sure that the $L$-th bond meets the $(L-1)$-th bond. Hence there are maximally $(2K-1)^L$ possible ``paths", and we can take the bound $(2K-1)^L<(2 K)^L$. Then we have,  by bounding $\norm{h}_\infty$ by $\abs{J}$,
\begin{widetext}
\begin{equation}\label{tb}
	\sum_{p:p_1\cap X\neq \emptyset,p_{|p|}\cap Z\neq \emptyset}  \frac{(|k_{\text{Max}} t|)^{|p|+1}}{(|p|+1)!} \left\|h(p)\right\|_\infty \left\|h_Z \right\|_\infty\leqslant
	\sum_{\abs{p}\geqslant L}  \frac{(2K|k_{\text{Max}} t J|)^{\abs{p}}}{\abs{p}!} 
\leqslant\sum_{\abs{p}\geqslant L}  \left(\frac{2eK k_{\text{Max}} |t J|}{\abs{p}}\right)^{\abs{p}}\leqslant y(t J,L)
\end{equation}
where $y(x,L)\equiv\text{exp}(-\mu_2 L)/(1-4Ke|x|/L)$ with $\mu_2=-\text{ln}(2Kek_{\text{Max}}| t J|/L)$. In Eq. \eqref{tb} we have used the bounds of Stirling's formula, i.e. $p!\geqslant \sqrt{2\pi}p^{p+1/2}e^{-p}\geqslant(p/e)^p$. 

Now combining  Eq.~\eqref{cff1}, Eq.~\eqref{nned} and Eq.~\eqref{tb}, we arrive at
\begin{equation}\label{nnbf}
	\norm{\text{Tr}([A(t), B] \rho(\beta))}_1 \leqslant \frac{4C\left\|A\right\|_\infty \left\|B\right\|_\infty}{\ln3(1-b(\beta J))(1-2Ke k_{\text{Max}} |t J|/L)}e^{-\mu_3 d(A,B)} 
\end{equation}
\begin{figure}[t]
\centering
\includegraphics[width=0.45\textwidth]{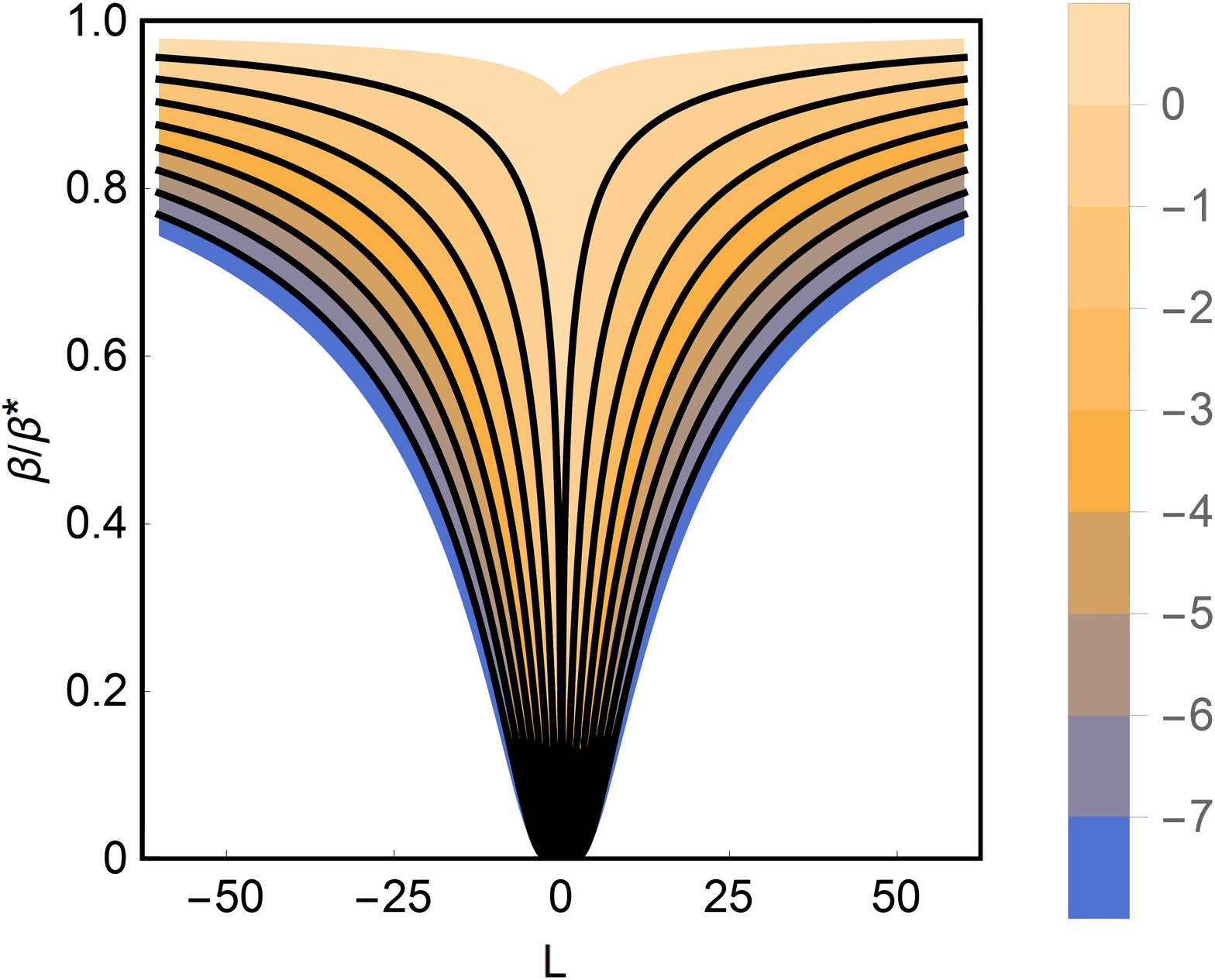}
\includegraphics[width=0.45\textwidth]{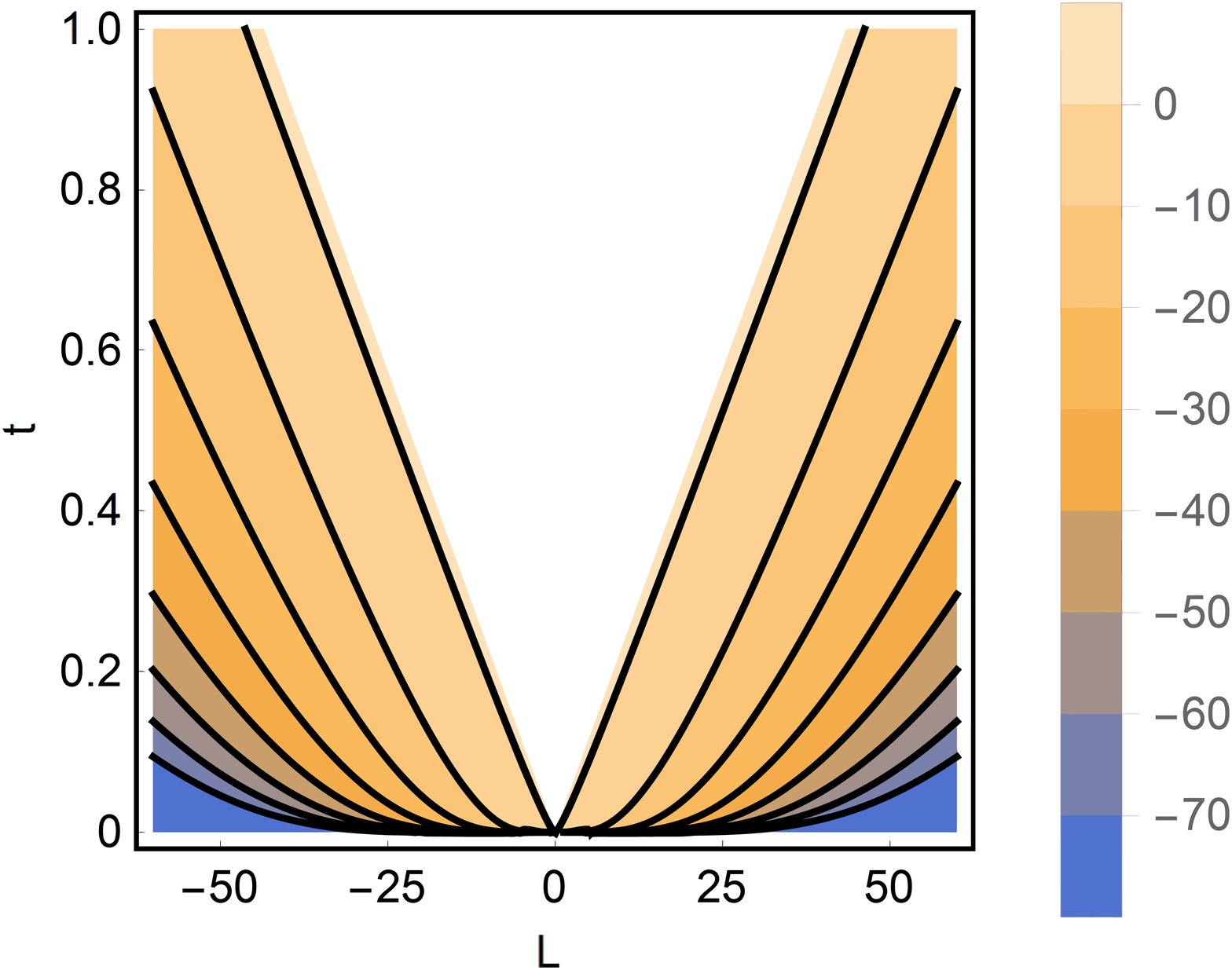}
\caption{The lightcone structure corresponding to \eqref{nnbf}. The left plot shows the bounds  at $t=0$. In such case the bound is a function of the inverse temperature $\beta$ in units of critical temperature and the width of the buffer region $L$. The right plot shows the bounds at $\beta=0$. We have chosen $K=4$, $J=1,k_{\text{Max}}=1$  and taken the  logarithm with base 10 of the bounds.}
\label{fig1}
\end{figure}
\end{widetext}
where $\mu_3=\text{min}\{\mu_1,\mu_2\}$ and $C$ is the lower bound of the number satisfying the inequality $\sum_{z\in V} e^{-\mu_3 (d(A,z)+d(z,B))}\leq C e^{-\mu_3 d(A,B)}$. Notice that the first term in \eqref{cff1} corresponds to the $t=0$ case and have been absorbed in the exponent in the second term of \eqref{cff1}. The lightcone structure now depends on the temperature (see Fig. \ref{fig1}).

Moreover, one can obtain a universal critical time from Eq.~(\ref{tb})
\begin{equation}
	t^*=\frac{L}{2 K e k_{\text{Max}}  | J|}.
\end{equation}
In spite of the temperature, if the evolution time $t$ is shorter than this critical time, according to Eq. \eqref{nnbf}, the correlation between two observables will be exponentially decaying with respect to the distance, so that the information cannot propagate to the distance $L$. Hence,
this critical time can be used to provide a universal lower bound on the Lieb-Robinson velocity in  many-body  systems.

\subsection{Systems with exponentially decaying interactions}\label{CCC}
For quantum lattice systems with exponentially decaying interactions, the interaction Hamiltonian $h_l$  decreases exponentially  with the  growth of  distance,
\begin{equation} \label{ied}
	\left\|h_l\right\|_\infty\leqslant \lambda_0 ~\text{exp}[-(\mu+\epsilon)d(x,y)]
\end{equation}
with some positive constants, $\lambda_0$, $\mu$ and $\epsilon$. We also require  the lattice to satisfy
\begin{equation} \label{sup}
	\sup_x \sum_{z\in V} \frac{1}{(1+d(x,z))^\epsilon}=\frac{p_0}{2^{\epsilon+1}}
\end{equation}
with a positive constant $p_0$. This equality can be used to derive another useful inequality
\begin{equation} \label{substract}
	 \sum_{z\in V} \frac{1}{(1+d(x,z))^\epsilon}\times\frac{1}{(1+d(z,y))^\epsilon}\leqslant\frac{p_0}{(1+d(x,y))^\epsilon}.
\end{equation} 
(Cf. the assumption 2.1. in \cite{HK06}.) 

\begin{figure}[t]
\centering
\includegraphics[width=0.5\textwidth]{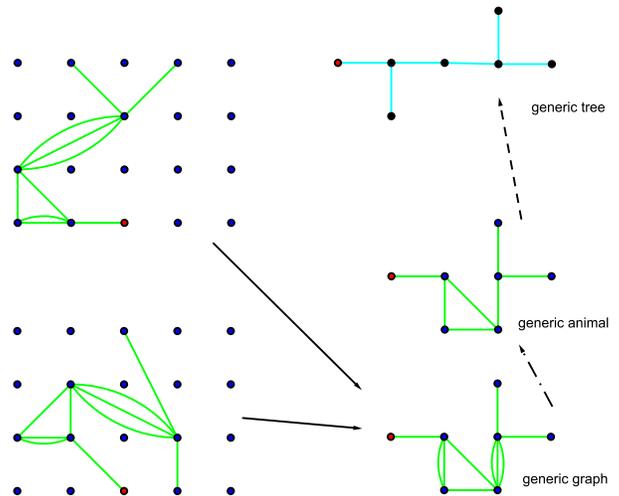}
\caption{The structure of generic graphs. When we choose the intermediate sites ergodically, the connective structures will not change. They belongs to the the same generic graph.}
\label{fig.e}
\end{figure}

Because the system might contain an infinite number of sites, and similarly an infinite number of adjacent bonds for each site. The method of generating function of graph  cannot be directly applied to this situation. To find a bound, we need to make use of Eq.~(\ref{substract}), which enables us  to contract two linked bonds into a single one. Particularly, let us define an equivalence relation between graphs by their connective structures of bonds (say, ordering), that is, two graphs are in the same {\it generic connective graph} if their connective structures are the same. (See also Fig. \ref{fig.e}) Given a generic connective graph starting with $x$ and ending on $y$, we can count the graphs in it by choosing all possible sites between $x$ and $y$ ergodically.
 Then the problem of counting all possible graphs (which could be infinite) is therefore translated into the problem of finding all possible generic connective graphs, which can be treated using the method introduced in Sec. \ref{II22}. Pointedly stated: (i) We divide generic connective graphs into connected parts. (ii) For the connected generic graphs, we  subtract the repeated bonds, and for the connected ones without repeated bonds, we further subtract the possible loops. (iii) Now the generic graphs are simplified to generic trees. We are going to analyze these tree graphs,  add the loops, reactivate repetitions,  glue the blocks, and go through all possible sites ergodically. Finally, we can recover all kinds of possible graphs thanks to the ergodic choices of intermediate sites. 

In a tree two neighboring bonds can always be contracted  and finally one is left with a longest single bond. For example, given  a tree $t$ connecting $x$ to $y$ with $k$ bonds, we can always find a generic tree $\mathfrak{t}$. We denote the set of (generic) trees connecting $x$ to $y$  by $\mathcal{T}$.
The ergodic choice of intermediate sites of $t$ can give all those trees corresponding to $\mathfrak{t}$. With  the help of Eq.~(\ref{ied}) and Eq.~(\ref{substract}), we can choose  the intermediate sites  ergodically through the lattice and obtain a bound on the interaction Hamiltonians defined on $\mathfrak{t}$,
\begin{equation}\label{tree}
	\sum_{t:t\to\mathcal{\mathfrak{t}}} \left\|h(t) \right\|_\infty\leqslant \frac{\lambda_0 ^k p_0^{k-1}\text{exp}(-\mu~ d(x,y))}{(1+ d(x,y))^\epsilon}.
\end{equation}

Next, in order to recover the loops and repetitions, we can fix one site $x$ and choose another site $y$ of $h_{xy}$ ergodically. Such a new graph may contain a loop, have a repeated  bond in the tree or become a larger tree. Applying Eq. \eqref{ied} and Eq.~\eqref{sup}, we obtain a new factor in the cluster expansion (of the partition function) from the above combinatoric considerations of adding a single bond,
\begin{widetext}
\begin{equation}
	\frac{k+1}{2} \sum_l\beta \left\|h_l\right\|_\infty\leqslant \frac{(k+1)\beta}{2}\sum_l \lambda_0 e^{-(\mu+\epsilon)d(x,y)}\leqslant \frac{(k+1)\beta\lambda_0}{2}\sum_l \frac{1}{(1+d(x,z))^\epsilon}
\leqslant\frac{(k+1)\beta \lambda_0 p_0}{2^{\epsilon +2}},
\end{equation}
where the factor $1/2$ comes from the indistinguishability of two sites of a bond $l$ and the sum over $l$ means that the intermediate sites are chosen ergodically. Recalling the example given in Sec. \ref{II22}, we see that the generating function of this combinatoric process is  $\mathcal{G}_{\text{gcg}}^k(x)=\exp\{\beta\lambda_0 p_0 (k+1)/{2^{\epsilon+2}}\}$, since adding an empty bond is allowed here just as adding nothing. 
We can in fact recover all possible loops and repetitions with this procedure.

Combining these with Eq.~\eqref{tree} and summing over all possible connected (generic) trees with different numbers of bonds, we can get a bound on  all possible connected graph $G$ in this situation. Consider first the term in the inclusion-exclusion form
\begin{align}
	\sum_{g\in G} \frac{(\beta)^{|g|}}{|g|!} \left\|h(g)  e^{\beta H_{\mathcal{T}(g)}}\right\|_\infty \leqslant&\sum_{t\in T} \left((\beta)^{|t|} \left\|h(t)\right\|_\infty \right) \left( \sum_{g'\in G'(t)} \frac{(\beta)^{|g'|}}{|g'|!} \left\|h(g')\right\|_\infty \right) \left\|e^{\beta H_{\mathcal{T}(t)}}\right\|_\infty\label{29} \\
	\leqslant& \sum_{t\in T} \left((\beta)^{|t|} \left\|h(t)\right\|_\infty \right)[\mathcal{G}_{\text{gcg}}^{|t|}(\beta\lambda_0)]^2\label{30}\\
\leqslant&\sum_{k=1}^\infty(4\beta)^{k}  \frac{\lambda_0 ^k p_0^{k-1}\text{exp}(-\mu~ d(x,y))}{(1+ d(x,y))^\epsilon} [\mathcal{G}_{\text{gcg}}^k(\beta\lambda_0)]^2\label{31}\\
 \leqslant&\frac{C_0}{(1+ d(x,y))^\epsilon}{ \left(\frac{f(\beta \lambda_0)}{1-f(\beta \lambda_0)}\right) \text{exp}(-\mu~ d(x,y))}, \label{cg}
\end{align}
where $f(x)\equiv4p_0x\text{exp} (2 x p_0 /{2^{\epsilon+2}})$ and $C_0=\text{exp} (2\beta\lambda_0 p_0 /{2^{\epsilon+2}}) p_0 ^{-1} $.  Here in the first bracket on the right hand side of \eqref{29} there is no redundency so that the generating function of a tree is simply $x^{\abs{t}}$ (cf. Sec. \ref{II22}), and the last two terms of \eqref{29} contribute to the $\mathcal{G}^2$ in \eqref{30} since they represent  the same process of adding  bonds in cases with and without existing bonds respectively. In \eqref{31} the factor $4^k$ is the bound of the number of possible plane trees with $k$ bonds. Indeed, since the connective structures of (generic) trees are determined, they are rooted ordered plane trees, and their numbers are counted by the $(k+1)$-th Catalan number (see, e.g. \cite{Kla97}). We therefore have for large number $n(=k+1)$ of sites, using Stirling's formula,
\begin{equation}
\frac{1}{n+1}\begin{pmatrix}2n\\n\end{pmatrix}=\frac{1}{n+1}\frac{(2n)!}{(n!)^2}\leqslant\frac{1}{n+1}\frac{e(2n)^{2n+1/2}}{2\pi n^{2n+1}}=\frac{e}{2\pi(n+1)}\frac{2^{2n+1/2}}{n^{1/2}}\leqslant4^n
\end{equation}
where the final inequality is chosen for the simplicity of including it into the $k$-th power in \eqref{31}.
 In \eqref{cg} the term in the bracket comes  from the sum of the $(4\beta)^{k}\lambda_0 ^k p_0^{k-1}[\mathcal{G}^k]^2$ over $k$, and of course the requirement $\abs{f}<1$ gives a critical inverse  temperature $\beta^*=f^{-1}(1)/\lambda_0$ similar to the $\beta^*$ in Sec.~(\ref{CCB}). With this, it is easy to give a bound to Eq.~\eqref{14},
\begin{equation}\label{33}
  \left\|h_Z B \rho(\beta)\right\|_1\leqslant 
  \left\|h_Z \right\|_\infty \left\|B \right\|_\infty\abs{Y} \sum_{m=1}^{\infty}  \frac{1}{m!}\prod_{i=1}^{m} \left[\left(\frac{f(\beta \lambda_0)}{1-f(\beta \lambda_0)}\right) \frac{C_0\text{exp}(-\mu~ d(Z,B))}{(1+ d(Z,B))^\epsilon}\right].
\end{equation}
Here $\abs{Y}$ is the number of sites in the support $Y$ of $B$. We can also use again the relation $ \text{exp}(x)-1\leqslant x(e^{x_0}-1)/x_0,\forall x\in[0,x_0]$ to simplify the bound \eqref{33}. For instance, at a high enough temperature such that $|Y| \left(\frac{f(\beta \lambda_0)}{1-f(\beta \lambda_0)}\right) C_0 \leq \text{ln}3 $, we have
\begin{equation}
  \left\|h_Z B \rho(\beta)\right\|_1\leqslant 
  \left\|h_Z \right\|_\infty \left\|B \right\|_\infty\abs{Y}  \left[ \left(\frac{f(\beta \lambda_0)}{1-f(\beta \lambda_0)}\right)  \frac{C_0 \text{exp}(-\mu~ d(Z,B))}{(1+ d(Z,B))^\epsilon}\right]\frac{2}{\text{ln}3}.
\end{equation}

The calculations about the ``paths" are relatively easier. We only need to contract the bonds using \eqref{substract} without the complications of generic connective graphs. The result is
\begin{equation}
	\sum_{p:p_1\cap X\neq \emptyset,p_{|p|}\cap Z\neq \emptyset}  \frac{(|k_{\text{Max}} t|)^{|p|+1}}{(|p|+1)!} \left\|h(p)\right\|_\infty \left\|h_Z \right\|_\infty\leqslant |X|p_0 ^{-1} e^{\lambda_0 p_0 k_{\text{Max}}  |t|} \frac{\text{exp}(-\mu~ d(A,Z))}{(1+ d(A,Z))^\epsilon}
\end{equation}
where $\abs{X}$ is the number of sites in the support $X$ of $A$.

Similar to Eq.~(\ref{nnbf}),  we finally have
\begin{equation}\label{nnbf1}
	{\abs{\text{Tr}([A(t), B] \rho(\beta))}} \leqslant \frac{4C_0 |X||Y|\left\|A\right\|_\infty \left\|B\right\|_\infty}{(1+ d(A,B))^\epsilon\ln3} \left(\frac{f(\beta \lambda_0)}{1-f(\beta \lambda_0)}\right)  {\exp^{-\mu~ d(A,B)+\lambda_0 p_0 k_{\text{Max}} |t|} }.
\end{equation}
\begin{figure}[t]
\centering
\includegraphics[width=0.45\textwidth]{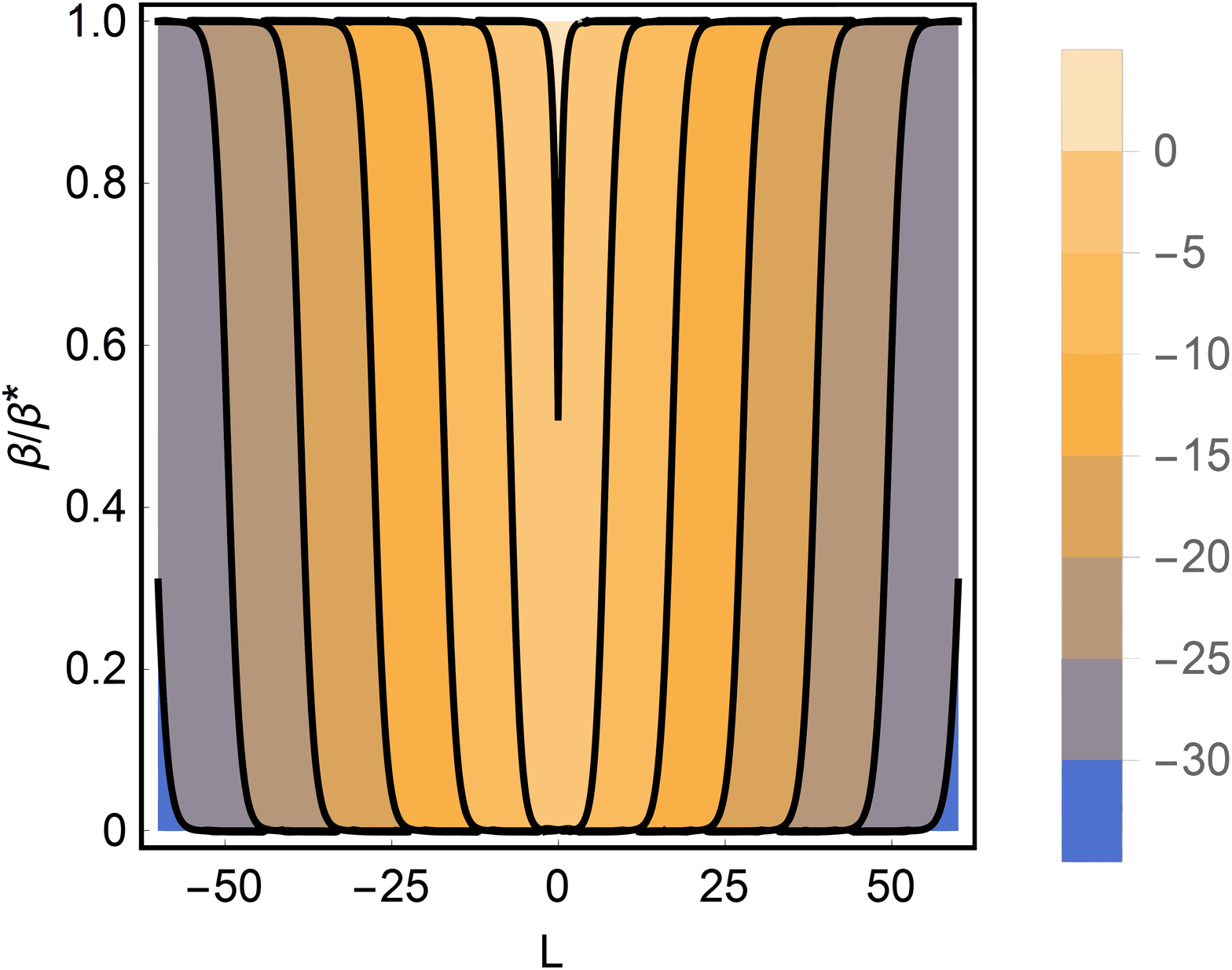}
\includegraphics[width=0.45\textwidth]{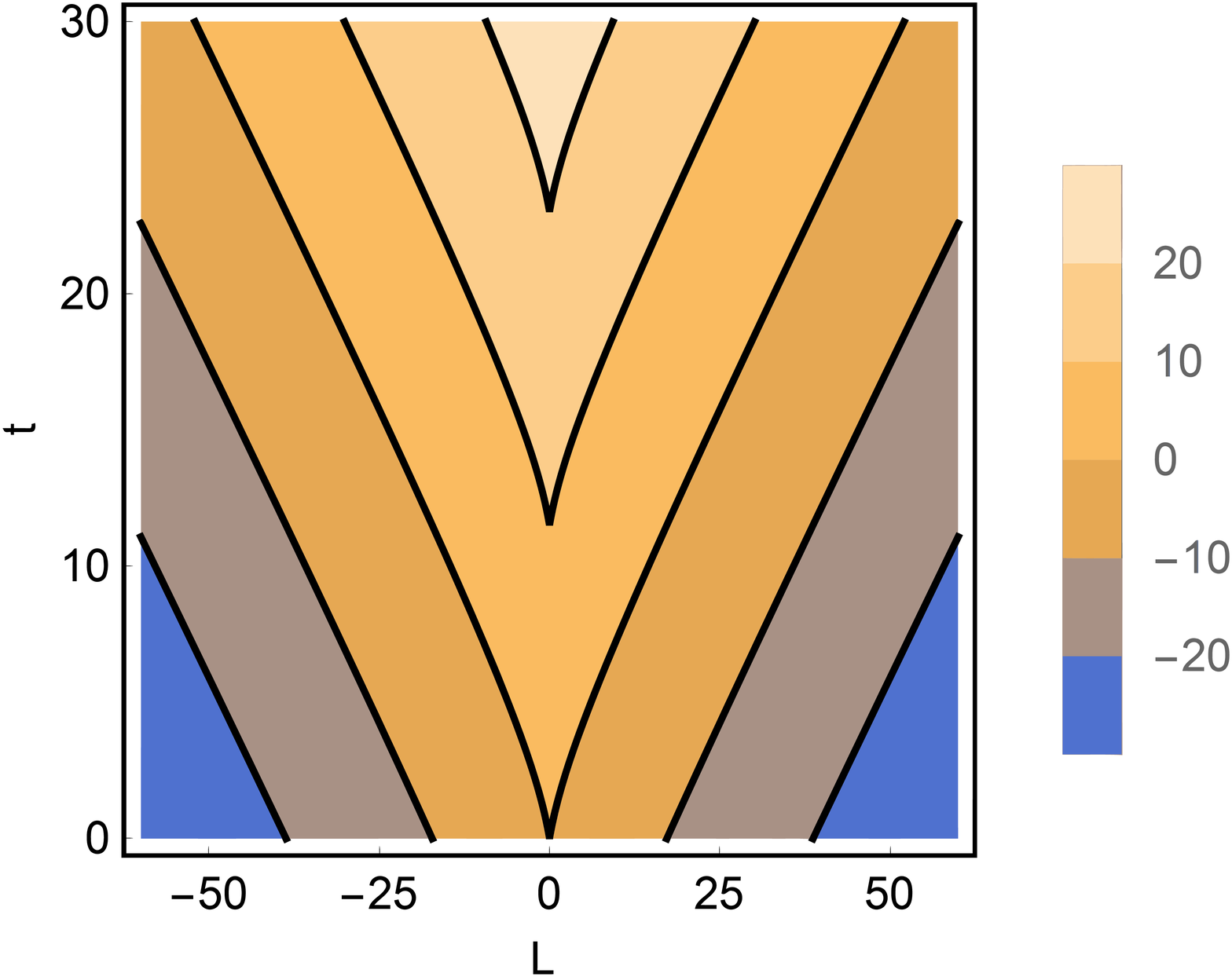}
\caption{The lightcone structure corresponding to \eqref{nnbf1}. The left plot shows the bounds at $t=0$, and the right plot shows the bounds  at $\beta=0$. Here $\epsilon=2$, $\mu=1$, $\lambda_0 p_0=1$, and we have taken $\log_{10}$ of the bounds.}
\label{fig2}
\end{figure}
\begin{figure}[t]
\centering
\includegraphics[width=0.45\textwidth]{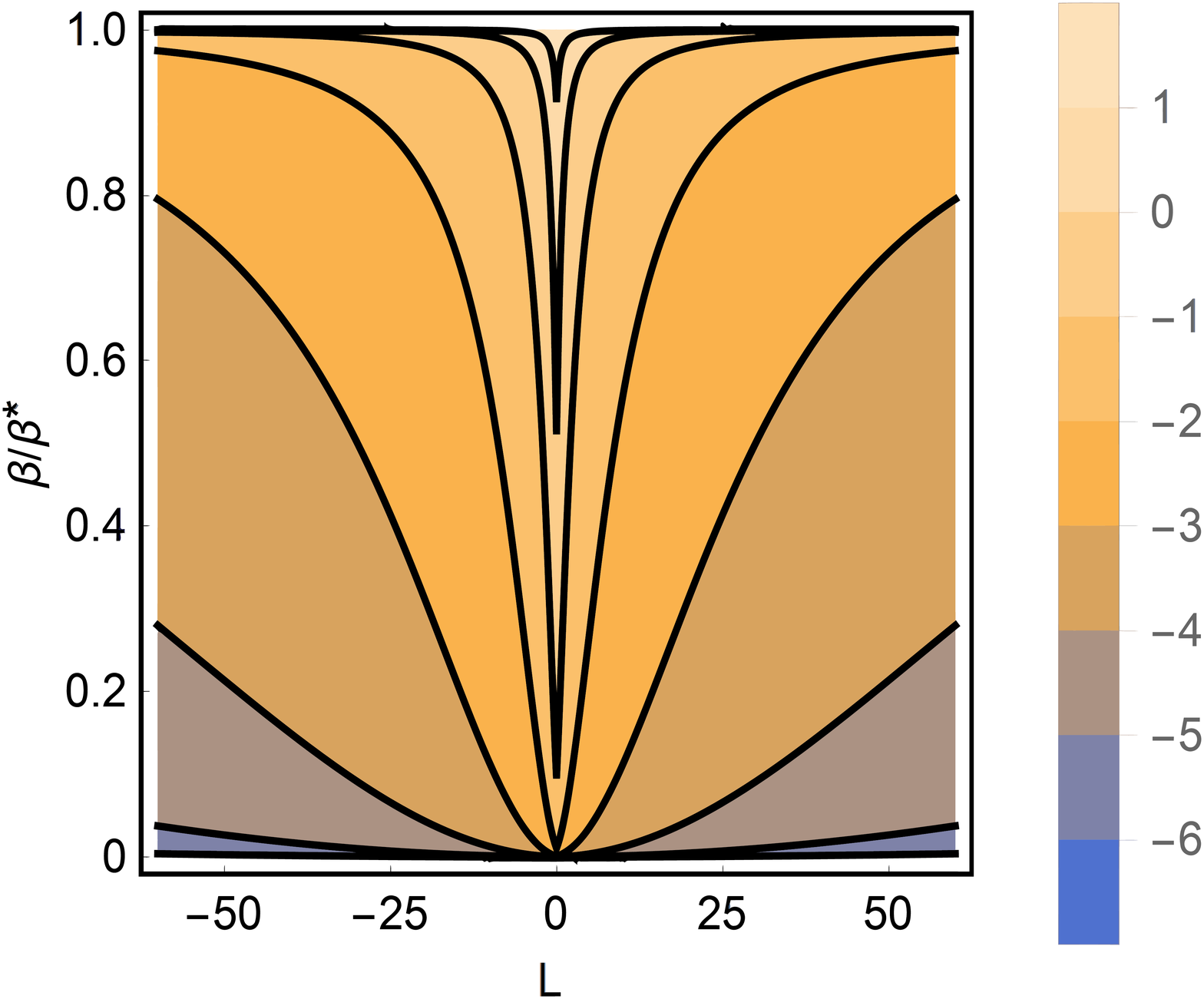}
\includegraphics[width=0.45\textwidth]{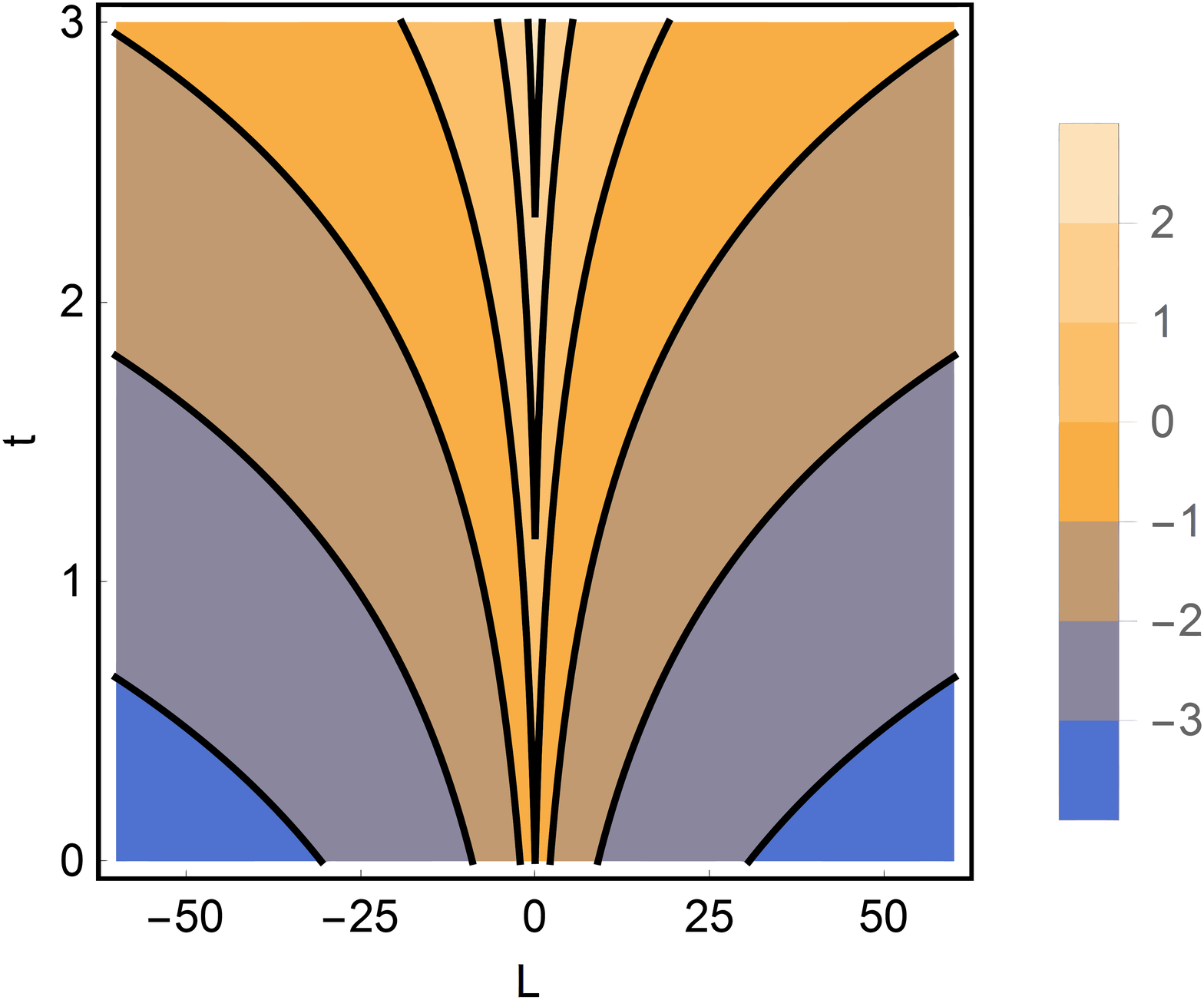}
\caption{The lightcone structure in systems with long-range interaction. The left plot shows the bounds at $t=0$, and the right shows the bounds at $\beta=0$.  Here $\epsilon=2$, $\mu=0$, $\lambda_0 p_0=1$, $k_{\text{Max}}=1$ and $\log_{10}$ has been taken.}
\label{fig3}
\end{figure}
\end{widetext}
The lightcone structure in this case is shown in Fig. \ref{fig2}.

\subsection{Systems with long-range interactions }
For the  systems with long-range interactions, $h_l$ is nontrivial for all of the bonds. But it decreases as power-law of distance
\begin{equation}
	\left\|h_l\right\|_\infty\leq \frac{\lambda_0}{(1+d(x,y))^\epsilon} .
\end{equation}
All the results of Sec.~(\ref{CCC}) can be applied here by setting the $\mu=0$. The lightcone structure is shown in Fig. \ref{fig3}.

By comparing the figures \ref{fig1}, \ref{fig2} and \ref{fig3}, we can clearly see how fast the interaction spread through the lattice, which is bounded by the Lieb-Robinson velocity.

\section{Conclusion and outlook}\label{dis}
In this paper we have proved the Lieb-Robinson-like bound at finite temperature for systems with respectively short-range, exponentially decaying and long-range interactions. We exploited the nontrivial technique of graph generating functions to count the number of ``paths" and ``clusters" in the cluster expansion, which greatly simplifies the proofs.

The temperature dependence of the Lieb-Robinson bound at finite temperature should have a wide range of applications, especially in comparing cases with different temperatures.  There are many other situations of direct relevance:
\begin{enumerate}
 \item The topological order has been proved to be stable under local perturbations with the help of the original Lieb-Robinson bound \cite{BHM10} or cluster expansions \cite{Kli10}. But the topological order does not exists at finite temperature  for many two-dimensional lattice models \cite{Has11}. From the Lieb-Robinson bound at finite temperature given in this paper,  a moment of reflection shows that there is a possible case where at a finite temperature the Lieb-Robinson velocity is still be relatively large, which implies the possibility  long-range entanglement at finite temperature. This is  corroborated by the result that the autocorrelation time of topological order at finite temperature, which is characterized by gaugelike symmetries \cite{NO08,NO09},  can be very large at a finite temperature below the spectral gap.
Given its importance in topological quantum memories at finite temperature, it is desirable if a more rigorous study of the dynamical thermal stability could be given.
\item The original Lieb-Robinson bound has recently been applied in holographic models, e.g. to bound the butterfly velocity \cite{RS16} and to bound the diffusivity \cite{HHM17}. We think an important missing point of these applications is that, due to the well-known Tolman-Ehrenfest effect \cite{TE}, in the gravitational dual bulk, say, a black hole spacetime with nonzero Hawking-Unruh temperature, even the equilibrium temperature cannot be the same constant, and not to mention those nonequilibrium effects listed above. So, the temperature  should have an impact of the localization properties in these cases. Our results have shown part of such impact.
\end{enumerate}

There is, however, a caveat that the obtained bound cannot be applied in quantum thermodynamics where the temperature dependence  is not needed. The original Lieb-Robinson bound can be directly applied because one starts with purely quantum setups and derive the emergent thermodynamic results. For recent applications, see e.g. \cite{IKS17,ST17}. Another drawback of the obtained bound is that the temperature effect and the time evolution are treated separately. In general, it is hoped that these two effects will affect each other, e.g. a Lieb-Robinson velocity depending on temperature. As is shown in \cite{Des13}, the Lieb-Robinson velocity in a dissipative quantum system is a function of the dissipation rate and will decrease due to the local dissipation. So, the question of how to relate the bound obtained above, which hold in the framework of quantum statistical mechanics, to the Lieb-Robinson bounds in stochastic dynamics deserves further investigations.

\begin{acknowledgments}
We thank Zohar Nussinov for helpful comments. We also thank the anonymous referees for their insightful comments on the earlier versions of the manuscript, especially on Eqs. \eqref{cff} and \eqref{cff1}.
Financial support from National Natural Science Foundation of China under Grant
Nos. 11725524, 61471356 and 11674089 is gratefully acknowledged.
\end{acknowledgments}

\end{document}